%% file: main.tex
\documentclass[conference,a4paper]{IEEEtran}
\usepackage{float}
\usepackage[utf8]{inputenc}
\usepackage{xcolor}
\usepackage[hidelinks]{hyperref}
\usepackage[numbers]{natbib}
\usepackage{lipsum}
\usepackage{graphicx}
\usepackage{amsmath}
\usepackage{cleveref}

\begin{document}

\title{Industrial Flexibility Investment Under Uncertainty: A Multi-Stage Stochastic Framework Considering Energy and Reserve Market Participation}
%Optimal Investment for Industrial Prosumers in Multi-Markets under Uncertainty: A Multi-Stage Stochastic Programming Framework
%Industrial Flexibility Investment Under Uncertainty: A Multi-Stage Stochastic Framework for Coordinated Participation in Energy and Reserve Markets

\author{Amund Norland$^a$, Lasse Skare$^a$, Ole Jakob Viken$^a$, \IEEEauthorblockN{Stian Backe$^{a,b*}$}
\IEEEauthorblockA{$^a$Department of Industrial Economics, Norwegian University of Science and Technology -- Trondheim, Norway\\
$^b$Energy Systems, SINTEF Energy Research -- Trondheim, Norway\\
%Email:\ stian.backe@ntnu.no
}}

\maketitle
\IEEEoverridecommandlockouts
\IEEEpubid{\makebox[\columnwidth]{979-8-3503-9042-1/24/\$31.00~\copyright~2025 IEEE \hfill} \hspace{\columnsep}\makebox[\columnwidth]{ }}
\IEEEpubidadjcol

\begin{abstract}
The global energy transition toward net-zero emissions by 2050 is expected to increase the share of variable renewable energy sources (VRES) in the energy mix. As a result, industrial actors will encounter more complex market conditions, characterized by volatile electricity prices, rising carbon costs, and stricter regulations. This situation calls for the industry to capitalize on opportunities in both spot-price arbitrage and reserve market participation, while also meeting future regulatory demands. This paper presents a multi-stage optimization framework that supports investment decisions in flexible assets and enables reserve market participation by delivering ancillary services. The framework incorporates investment decisions, spot- and reserve-market bidding, and real-time operation.  Uncertainty in market prices and operational conditions is handled through a nodal formulation. A case study of a large industrial site in Norway is performed, comparing the investment decisions with future technology- and carbon pricing scenarios under varying market conditions.

\end{abstract}

\input{Seksjoner/01_Intro}
\input{Seksjoner/01B_Modeling_framework}
\input{Seksjoner/02_Mathematical_formulation}
\input{Seksjoner/03_case_study}

\input{Seksjoner/04_Results_and_conclusion}

% \section*{Acknowledgment}
% This paper was prepared as a part of the FLXenabler project funded by the Research Council of Norway and the U.S. Department of Energy through the ERA-NET program No. 341596. 
%\bibliographystyle{IEEEtranN}
%\bibliography{LinkRef}
\input{main.bbl}

\end{document}

%% file: Seksjoner/01_Intro.tex
\section{Introduction} \label{sec:intro}
The increasing share of VRES in the energy mix is expected to transform the electricity markets, presenting both opportunities and challenges for industrial prosumers. To enhance grid stability, market operators have introduced systems/markets that ensure sufficient capacity is available to manage future fluctuations. These systems are divided into frequency control - responsible for balancing supply and demand to maintain 50 Hz - and non-frequency control, which focuses on voltage control, local grid stability, and black-start capabilities. For this paper, only frequency control reserves are considered.

Historically, grid stability has been maintained by power producers. However, as the share of intermittent renewables grows, the Transmission System Operator (TSO) face increasing difficulties in preserving system balance. In response, TSOs have established several markets, each tailored to specific types of frequency deviations. A common feature of these markets is that they offer participants - such as industrial prosumers - an opportunity to earn additional revenue by providing flexibility and ancillary services.

The industrial sector, accounting for around 40\% of the global energy consumption, offers a unique opportunity by enhancing this kind of market flexibility. Many heavy industries have the ability to switch between energy carriers, and their substantial electricity demand allows them to meet the minimum volume thresholds required for participation in reserve market through e.g. load shifting, in contrast to smaller consumers.

Extensive research has explored multi-market participation under uncertainty, focusing on both coordinated and independent bidding strategies. \citet{klaeboe2013optimal} found that coordinated bidding yields modest benefits, with gains between 0.1\% and 2\%, and suggested that these may increase as price volatility grows. In contrast, \citet{lohndorf2023value} reported significantly higher gains — ranging from 8\% to 29\% —when coordinated bidding involves flexible storage systems across integrated day-ahead, intraday, and reserve markets. Moreover, \citet{boomsma2014bidding}, \citet{ottesen2018multi}, \citet{bohringer2019trading}, \citet{bohlayer2020energy} and \citet{Paredes_multi_market} have extended these frameworks by combining different capacity and activation participation within reserve markets. Their studies demonstrate overall profitability of market participation; nevertheless, most of the research is conducted from the perspective of aggregators or Balancing Responsible Parties (BRP). 

Research on industrial flexibility and demand-side management (DSM) has identified promising technologies, ranging from large-scale heat pumps and Battery Energy Storage Systems (BESS) to Thermal Energy Storage (TES). Common for them all, is that they enable larger industries to mitigate grid fluctuations and capitalize on volatile market prices through energy carrier switching or spot-arbitrage \citep{mathiesen2009comparative_an_seven_tech, beck2021optimal_selection_of_storage}. 

Although several operational research frameworks has been developed, few have addressed the integration of investment decisions and bidding strategies for spot, intraday, and reserve markets from an industry-specific perspective within a multi-stage stochastic optimization framework. In addition, there is limited research on how uncertainty could affect future investment decisions. This paper builds on existing research by incorporating the following features:
\begin{itemize}
    \item Storage and conversion technology investment decisions
    \item Industry specific perspective
    \item Intraday bidding
    \item A nodal formulation for uncertainty handling
\end{itemize}

The identified research gap, combined with the complexities of market design and regulatory requirements, along with uncertainties in future price and market conditions, motivates the following problem statement:

\begin{itemize}
    \item How do variations in uncertainty representation and future market conditions - such as reserve market participation, increasing emission costs, and declining investment cost - affect investment decisions in flexible assets for industrial prosumers in a multi-market approach.
\end{itemize}

%\citet{connolly2017heat} find that this is important \cite{bloess2018power}

%% file: Seksjoner/01B_Modeling_framework.tex
\section{Modeling Framework} \label{sec:Modeling_framework}

\begin{figure}[h]
    \centering
    \includegraphics[width=\linewidth]{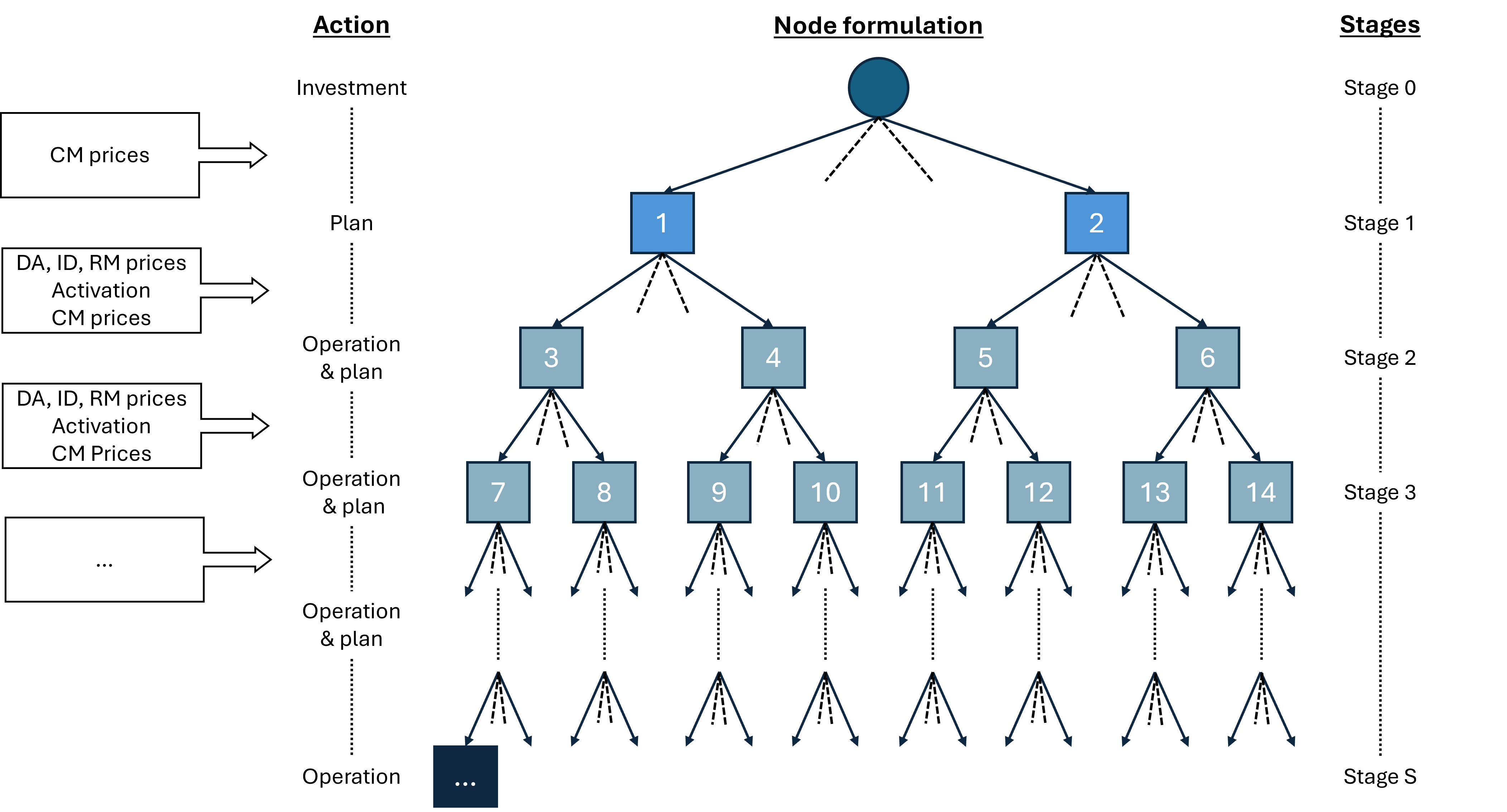}
    \caption[Modeling framework logic]{Modeling framework logic, visualizing  how uncertainty is revealed and represented within a nodal framework and how actions, stages and operational periods are represented. (Abbreviations: CM = Capacity market, DA = Day-ahead, ID = Intraday, RM = Reserve markets)}
    \label{fig:framework_representation}
\end{figure}

To properly address the problem statement, a multi-stage stochastic modeling framework is developed. The logic of this framework is illustrated in \autoref{fig:framework_representation}, which presents the node-based formulation of the model and highlights how uncertainty is progressively revealed and how actions, stages  and operational periods are represented throughout the model. The model is generalized, allowing a user-defined number of stages, time steps, and node-branches in each stage to represent various setups and uncertainty representations.

The market scope includes bids in the capacity, day-ahead and intraday markets, as well as real-time operation and activation. Seasonal bidding and stand-alone activation markets are excluded from this study and left for future work. The framework includes only one single reserve market, based on the assumption that all relevant markets have been merged into one. Alternatively, a preliminary analysis can be performed to identify the most relevant market prior to constructing the model instances. Additionally it is assumed that all flexible storage technologies modeled are qualified to participate in the chosen market. A more detailed analysis involving multiple reserve markets and technology-specific qualification criteria is left for future work.

The framework’s stage-wise decision structure and uncertainty realization are developed based on conversations with several Norwegian industrial actors, providing insight into their internal processes and deadlines for market and operational decisions. Although the market sequence (day-ahead, intraday, and reserve) has clearly defined, sequential gate closures, these discussions revealed that industrial actors often follow internal deadlines imposed by their BRP or aggregator. As a result, they typically submit bids for all markets simultaneously, without the opportunity to revise decisions based on updated information or uncertainty realizations between market stages. Consequently, this paper consolidates the day-ahead, intraday, and capacity market bidding decisions into a single decision stage, as illustrated in \autoref{fig:framework_representation}.

%% file: Seksjoner/02_Mathematical_formulation.tex
\section{Mathematical formulation} \label{sec:Mathematical_formulation}

The primary objective is to support investment decisions while enhancing flexibility potential and enabling participation in multiple energy markets under uncertainty. Taking into account the site's specific energy demands, load-shifting constraints, available on-site technologies, and flexible assets, the framework proposes a cost-minimizing strategy for both investment and operation. This is achieved by explicitly incorporating uncertainty related to obligations in the day-ahead, intraday, and reserve markets, alongside operational decision-making, as initiated in the objective function (\ref{eq:objective}). $t$ represents the time step, $s$ the stage, $\omega$ the node, and $\pi_\omega$ is the probability of node $\omega$, where $\sum_{\omega \in \Omega_s}\pi_\omega = 1 \quad \forall s \in \mathcal{S}$. 

%which include operation and maintenance (O&M) costs for storage technologies and the inconvenience costs associated with downshifting load.

The first-stage decision allows for capacity expansion ($I^\text{inv}$) on-site if found economically reasonable, considering exogenously defined normalized investment costs and an upper budget limitation. The next stages, depending on the amount of operational periods, includes revenues from the capacity reserve market ($I_t^\text{CM}$), cost or revenues from reserve activation ($I_{\omega t}^\text{ACT}$), day-ahead ($I_{\omega t}^\text{DA}$) and intraday market positions ($I_{wt}^\text{ID}$), as well as operational expenditures ($I_{wt}^\text{OPEX}$), which include O\&M costs and degradation for storage technologies and the inconvenience costs associated with load shifting. For the grid tariff, $I^\text{GT}$, only the power component is included. The mathematical model reflects the framework represented in \autoref{fig:framework_representation}.
\begin{align}
    \min z = I^\text{inv} &+ I^\text{GT} + \sum_{t \in \mathcal{T}} \left( I_t^\text{CM} + \sum_{s \in \mathcal{S}}\sum_{\omega \in \Omega_s} \pi_\omega \left( I_{\omega t}^\text{ACT} \right.\right.\nonumber \\ & \left.\left.+ I_{\omega t}^\text{DA} +I_{wt}^\text{ID} +  I_{wt}^\text{OPEX}\right)\right)
    \label{eq:objective}
\end{align}

The energy balance in Constraint \ref{eq:energy_balance} ensures that the system's demand and supply are fulfilled across all time steps $t$, nodes $\omega$ and energy carriers $e$. The demand is represented by the reference demand $D^\text{ref}$, which can be adjusted through load shifting. Here, $\Delta^\text{LS$\uparrow$}$ indicates upward shifted demand and $\Delta^\text{LS$\downarrow$}$ indicates downward shifted demand, relative to the baseline. $y^\text{out}$ and $y^\text{in}$ represent the energy flow from technology to fuel, and fuel to technology, respectively. \autoref{fig:Conversion_matrix} provides an illustration of the overall system interactions.
\begin{flalign}
    D_{\omega te}^\text{ref} + \Delta_{\omega te}^\text{LS$\uparrow$} - \Delta_{\omega te}^\text{LS$\downarrow$} &= \sum_{o \in \mathcal{O}}\left(\sum_{i \in \mathcal{I}_{eo}^\text{out}}  y_{\omega tieo}^{\text{out}} - \sum_{i \in \mathcal{I}_{eo}^\text{in}} y_{\omega tieo}^{\text{in}}\right) \nonumber \\ &\quad- \sum_{b \in \mathcal{B}_e} ( \eta_{b}^{\text{charge}} q_{\omega tb}^{\text{charge}} -  q_{\omega tb}^{\text{discharge}})  \nonumber \\ &\quad \forall s \in \mathcal{S}, \omega \in \Omega_s, t \in \mathcal{T},  e \in \mathcal{E} &
    \label{eq:energy_balance}
\end{flalign}

The market balances ensure that the traded electricity does not exceed what's available on-site  or through market trading.  These constraints apply specifically to the power grid (PG), the electricity energy carrier (EL), and modes ($o$) for import and export. Intraday trading volumes are further limited by a fraction of historical average volumes, reflecting lower market liquidity relative to the day-ahead market. Constraint \ref{eq:Market_balance_out} illustrates the import balance, while an equivalent exists for exports.
\begin{flalign}
    y_{\omega tieo}^{\text{out}} &= x_{pt}^{\text{DA,buy}} + \sigma_{\omega t}^\text{ID,buy}x_{pt}^\text{ID,buy} + \sigma_{\omega t}^\text{DWN} x_{pt}^\text{DWN} \nonumber \\  & \quad\quad\quad\quad\quad\quad\quad\quad\forall s \in \mathcal{S}, \omega \in \Omega_s, p \in \mathcal{P}_\omega,  t \in \mathcal{T},\nonumber \\& \quad\quad\quad\quad\quad\quad\quad\quad i = \text{PG}, e=\text{EL}, o=1 &
    \label{eq:Market_balance_out}
\end{flalign}

The level of activity of each technology $i$ is physically connected to the supply of corresponding energy carriers through efficiencies ($\eta$). Constraint \ref{eq:Tech_To_EC_conversion} illustrates this relationship for energy outputs, with a similar constraint existing for energy inputs. 
\begin{flalign}
    y_{\omega tieo}^{\text{out}} &= y_{\omega tio}^\text{activity} \eta^{\text{out}}_{ieo} \nonumber \\ &\quad \forall  s\in \mathcal{S}, \omega \in \Omega_s, t \in \mathcal{T}, e \in \mathcal{E}, o \in \mathcal{O}, i \in \mathcal{I}_{eo}^\text{out}&
    \label{eq:Tech_To_EC_conversion}
\end{flalign}

To reflect real-world limitations in the operation of technologies, ramping constraints are included to limit the rate of change in technology output between consecutive time steps. Constraint \ref{eq:Tech_ramping} ensures this within each time step, using the ramping factor, $\beta_i$, based on both installed (\( V^{\text{init}}_{i} \)) and added capacity (\( v_{i}^{\text{new}} \)). Additional constraints exist to handle ramping across stages and for the first time step, but are omitted here for brevity.
\begin{flalign}
    &y_{\omega tieo}^{\text{out}} - y_{\omega ,t-1,i,e,o}^{\text{out}} \leq \beta_i \left(V^{\text{init}}_{i} +v_{i}^{\text{new}} \right) \nonumber \\ &\quad \forall s\in\mathcal{S}, \omega \in \Omega_s, t \in \mathcal{T} \setminus \{1\}, e \in \mathcal{E}, o \in \mathcal{O}, i \in \mathcal{I}_{eo}^\text{out}&
    \label{eq:Tech_ramping}
\end{flalign}

In order to limit the heat pump to real-world practices, the output has to be restricted by the available excess heat at site and the supplied electricity. Constraint \ref{eq:Heat_pump_input_limitation} captures the available excess heat across all heat demands, and allocates the heat input among the different heat pump technologies. It also allows heat to be supplied to multiple temperature levels, under the condition that higher-temperature heat pumps can serve lower-temperature demands, but not vice versa.
\begin{flalign}
    &\sum_{i \in \mathcal{I}^\text{HP}} \sum_{o \in \mathcal{O}_i} \left(\sum_{e \in \mathcal{E}^\text{Heat}} y_{\omega tieo}^\text{out} - y_{\omega tie'o}^\text{in}\right) \leq \mu^\text{surplus} \sum_{e\in \mathcal{E}^\text{Heat}}d_{\omega te}^\text{flex}& \nonumber \\ & \hfill\quad\quad\quad\quad\quad\quad\quad\quad\quad\forall s \in \mathcal{S}, \omega \in \Omega_s, t \in \mathcal{T}, e' = \text{EL}&
\label{eq:Heat_pump_input_limitation}
\end{flalign}

To model temporal demand-side flexibility without curtailing load, the model allows for upward and downward load shifting while ensuring that the total energy demand is met over the given planning horizon. Constraint \ref{eq:LS_AGG_BALANCE} ensures that aggregated upward and downward shifts are balanced in selected periods ($s \in \mathcal{S}^\text{shift}$). The magnitude of shifted load is limited to a fraction of the reference demand. Additional constraints ensuring initialization, aggregation, and time steps where load shifting is permitted are included in the model but omitted here for brevity.
\begin{flalign}
    \text{LS}_{\omega e}^\text{Agg$\uparrow$} &= \text{LS}_{\omega e}^\text{Agg$\downarrow$} & \forall s \in \mathcal{S}^\text{shift}, \omega \in \Omega_s, e \in \mathcal{E}\quad
    \label{eq:LS_AGG_BALANCE}
\end{flalign}

Constraint \ref{eq:DWN_reserve_limit} sets an upper bound for the downward reserve capacity. This is proposed limited by the maximum bid size of indivisible bids, as bids are assumed activated in their entirety. Additionally, the model includes constraints that restrict bid sizes to a fraction of historically procured volumes, providing an alternative upper bound when stricter than Constraint \ref{eq:DWN_reserve_limit}. Similar constraints apply to upward reserves, but are omitted here for brevity.
\begin{flalign}
    & x_{\omega t}^\text{DWN} \leq X^\text{Max} & \forall \omega \in \Omega, t \in \mathcal{T}\quad&
    \label{eq:DWN_reserve_limit}
 \end{flalign}
 
Constraint \ref{eq:battery_charge_discharge_limit} ensures that the sum of energy charged and discharged at time $t$, adjusted for efficiency, does not exceed the maximum capacity of flexible asset $b$. $\theta_b$ represents the power-to-energy ratio, defined as the ratio between the installed capacity, and the corresponding power output. This formulation limits simultaneous charging and discharging, which, in practice, is meaningless.
\begin{flalign}
 &q_{\omega tb}^{\text{charge}} + \frac{q_{\omega tb}^{\text{discharge}}}{\eta_b^\text{discharge}}  \leq P_{b}^{\text{max}} + \theta_b v_b^\text{new} \nonumber \\  & \quad\quad\quad\quad\quad\quad\quad\forall s \in \mathcal{S}, \omega \in \Omega_s, t \in \mathcal{T}, e \in \mathcal{E}, b \in \mathcal{B}_e&
\label{eq:battery_charge_discharge_limit}
\end{flalign}

The state of charge (SoC) of flexible asset $b$ is determined by the energy stored in the previous time step, as well as the amounts charged and discharged while accounting for storage losses (\(\varepsilon_{b}^{\text{loss}}\)), as defined in Constraint \ref{eq:battery_storage_dynamics}. Additional constraints handling continuity across stages, initialization, end-of-horizon, and maximum storage capacity exist, but are excluded in this paper for brevity.
\begin{flalign}
&q^\text{SoC}_{\omega tb} = q^\text{SoC}_{\omega,t-1,b}(1 - \varepsilon_{b}^{\text{loss}}) + {q_{\omega tb}^{\text{charge}}} -  \frac{ q_{\omega tb}^{\text{discharge}}}{\eta_{b}^{\text{discharge}}}  \nonumber \\ &\quad\quad\quad\quad \forall  s \in \mathcal{S}, \omega \in \Omega_s, t \in \mathcal{T} \setminus \{1\}, e \in \mathcal{E}, b \in \mathcal{B}_e&
\label{eq:battery_storage_dynamics}
\end{flalign}

The energy supplied by technology \(i\) summed over all energy carriers \(e\) and mode of operation $o$ is limited by the sum of the initial installed capacity and the added capacity, multiplied by the availability factor $\phi_{\omega ti}$. 
\begin{flalign}
     \sum_{o \in \mathcal{O}_i} \sum_{e \in \mathcal{E}_i^\text{out}} y_{\omega tieo}^{\text{out}} &\leq \phi_{\omega ti} \left(V^{\text{init}}_{i} +v_{i}^{\text{new}} \right)   \nonumber \\ &\quad\quad\quad \forall s \in \mathcal{S}, \omega \in \Omega_s, t \in \mathcal{T}, i \in \mathcal{I}&
\label{eq:Availability_limitation}
\end{flalign}

Constraint \ref{eq:Plusskunde_export_limitation} establishes an upper bound on electricity export, proposed based on Norwegian prosumer regulations, where exceeding this limit incurs additional fees.
\begin{flalign}
    &y_{\omega tieo}^{\text{in}} \leq G^\text{export}  & \forall s \in \mathcal{S}, \omega \in \Omega_s, t \in \mathcal{T}, e = \text{EL}, o=2 
\label{eq:Plusskunde_export_limitation}
\end{flalign}

To represent the power component used in the grid tariff calculation, the model identifies the peak grid import within each month $m$. Constraint \ref{eq:peak_power_for_grid_tariff} captures the highest grid import ($y^\text{out}$) at each node $\omega$, and assigns it to a peak variable $y^\text{max}$. A supporting constraint, omitted here, ensures this peak value is propagated from parent to corresponding child nodes such that the final stage of each month reflects the correct peak load for tariff calculation.
\begin{flalign}
    &y_{\omega tieo}^\text{out} \leq y_{\omega}^\text{max}  \nonumber \\ & \quad \forall m \in \mathcal{M}, \omega \in \Omega_m, t \in \mathcal{T}, i=\text{PG},e=\text{EL}, o \in \mathcal{O}&
\label{eq:peak_power_for_grid_tariff}
\end{flalign}

Although emission costs are included in the objective, additional emission limits may still be relevant due to regulatory or company-specific requirements. Constraint \ref{eq:Emission_limit} imposes an upper bound on emissions, with $\gamma_{io}$ representing the carbon intensity of technology $i$ in operational mode $o$, including auxiliaries. 
\begin{flalign}
    &\sum_{o \in \mathcal{O}}\sum_{i \in \mathcal{I}}\sum_{t \in \mathcal{T}}y_{\omega tio}^\text{activity} \gamma_{io} \leq \gamma^{max} & \forall s \in \mathcal{S}, \omega \in \Omega_s\quad
    \label{eq:Emission_limit}
\end{flalign}

%% file: Seksjoner/03_case_study.tex
\section{Experimental setup and case study}\label{sec:case_study}
\begin{figure}[htp]
    \centering
    \includegraphics[width=\linewidth]{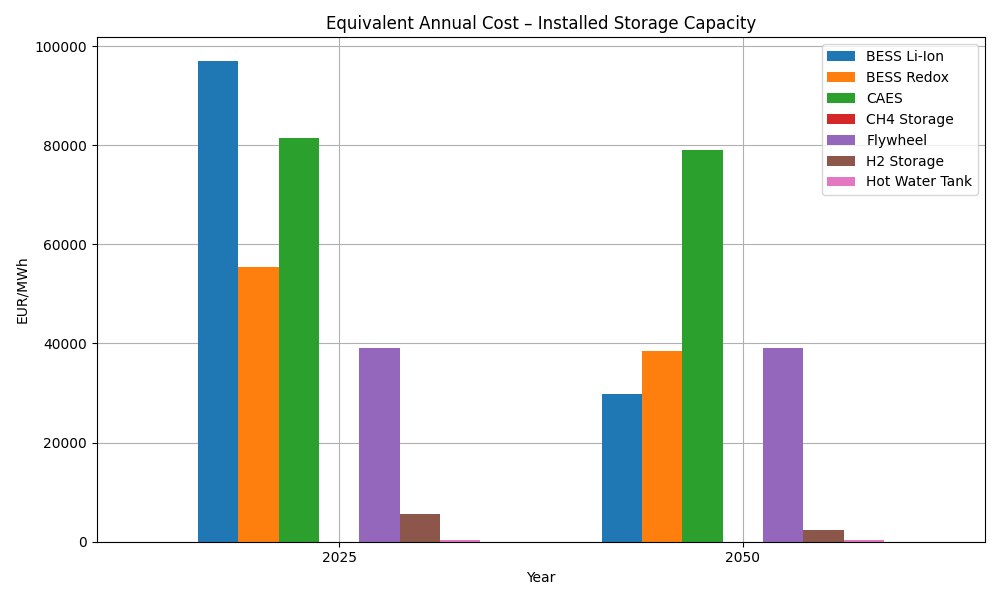}
    \caption{Equivalent annual cost of storage, using a discount rate of 9\%}
    \label{fig:CAPEX_storage}
\end{figure}

\begin{figure}[htp]
    \centering
    \includegraphics[width=\linewidth]{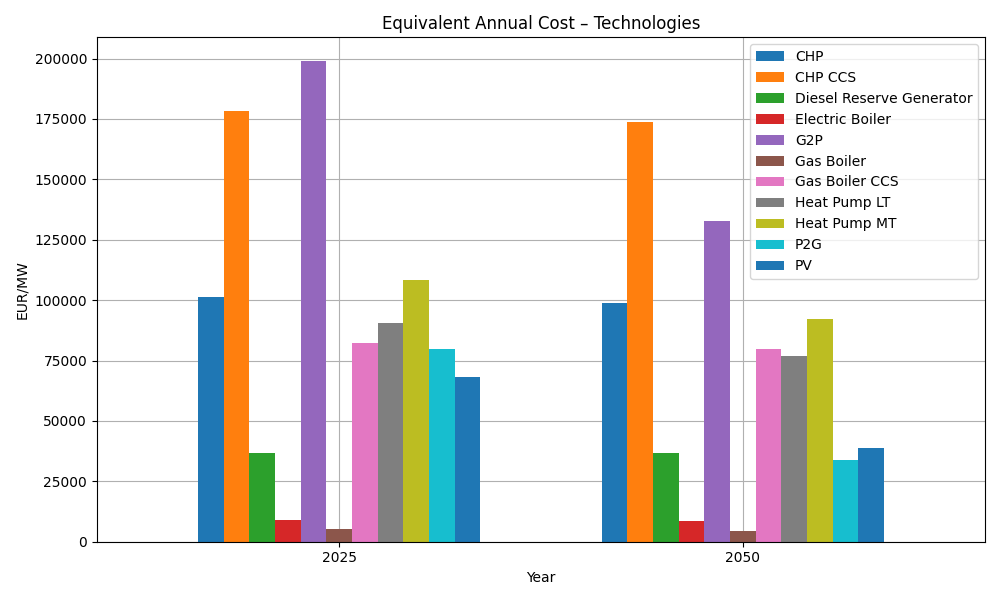}
    \caption{Equivalent annual cost of technologies, using a discount rate of 9\%}
    \label{fig:CAPEX_conversion_tech}
\end{figure}

The experimental setup is based on relevant data obtained from conversations with Norwegian industrial actors, including representative load curves for heat and electricity. The storage and conversion technologies considered, along with their investment costs, are illustrated in \cref{fig:CAPEX_storage,fig:CAPEX_conversion_tech}, using technology-specific data from the Danish Energy Agency. Investment costs are normalized over the planning horizon using a 9\% discount rate. Electricity prices, reserve market prices, and solar generation profiles are based on historical data from price zone NO1, with mFRR serving as the selected reserve market.

The model is implemented in Python using Pyomo, with Gurobi as the solver. Each stage consists of 24 time steps, representing one day. Rather than modeling a long time horizon with fewer branches, the case study only includes one operational day with a larger number of branches to better capture uncertainty realizations. In order to ensure that the investment decisions are representative for seasonal variability, the first-stage branches are clustered into four representative scenarios, each capturing distinct seasonal demand, prices and market characteristics. 

\begin{figure}[htp]
    \centering
    \includegraphics[width=0.9\linewidth]{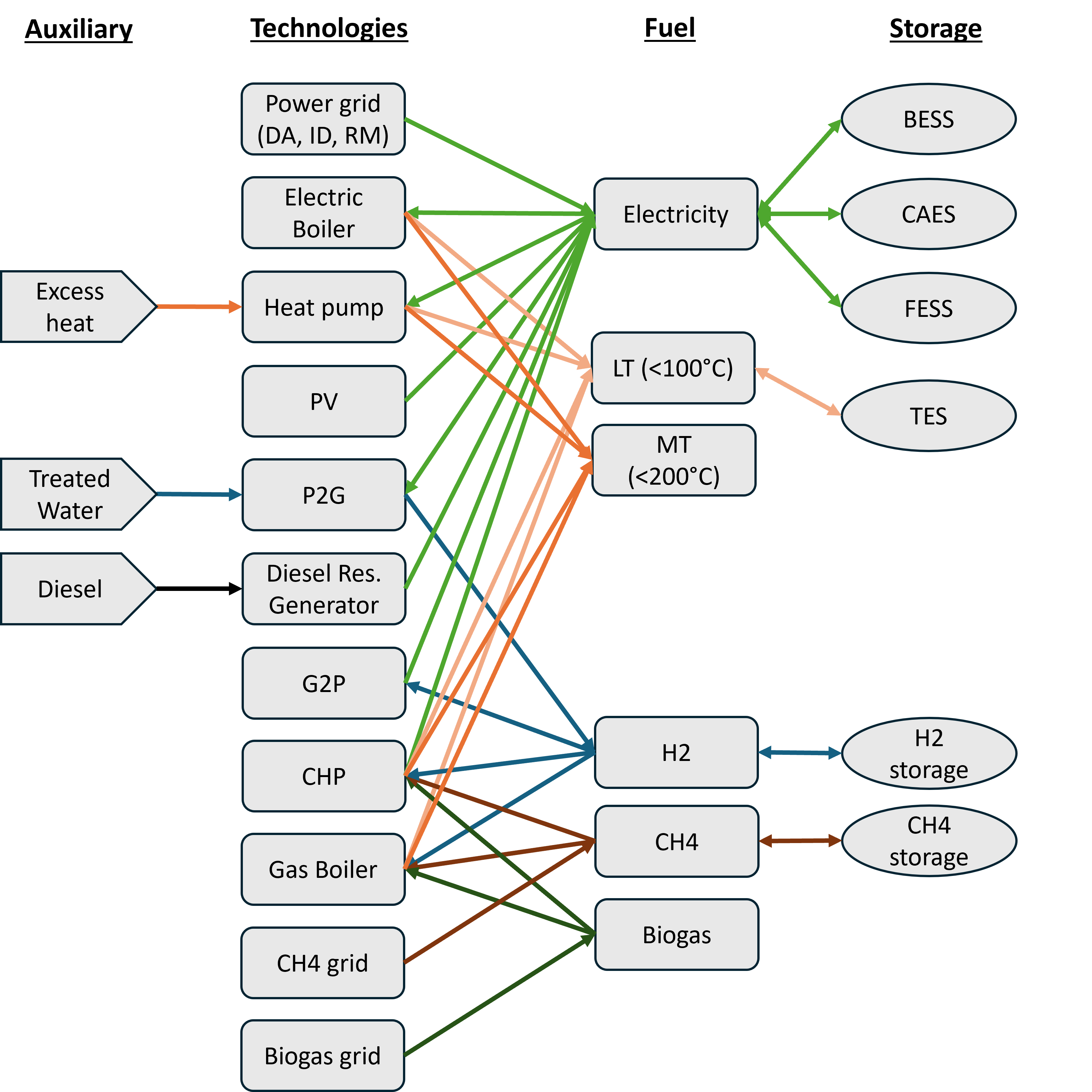}
    \caption{Possible interaction between auxiliaries, technologies, fuels and storage in the case study.}
    \label{fig:Conversion_matrix}
\end{figure}

\autoref{fig:Conversion_matrix} illustrates the connection between fuels, technologies, and storage options. The demand for electricity and heat is defined exogenously, while the others are dummy-fuels to enable multi-fuel flexibility. The case study assumes no pre-installed capacity aside from the grid connections, allowing the model to fully determine the optimal technology mix among the available technologies.

Load shift up to 10\% above and 30\% below the reference demand $D_{\omega te}^\text{ref}$ is allowed during time steps $\left[\text{8-17}\right]$, provided that the total reference demand is met over the planning period.

%% file: Seksjoner/04_Results_and_conclusion.tex
\section{Results and discussion} \label{sec:results}
\begin{figure}[htp]
    \centering
    \includegraphics[width=0.9\linewidth]{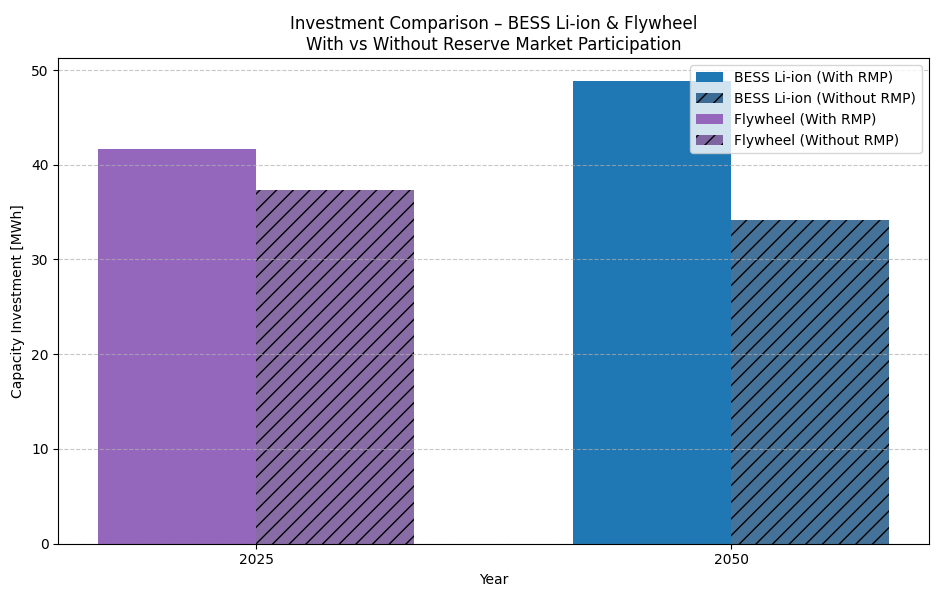}
    \caption{Storage technology investments in 2025 and 2050 with and without reserve market participation (RMP).}
    \label{fig:storage_investments}
\end{figure}

\autoref{fig:storage_investments} illustrates how battery storage investments differ across scenarios for 2025 and 2050, both with and without reserve market participation. As depicted, participation in the reserve market has a significant impact on the optimal sizing of storage technologies. In 2025, flywheel is favored, while by 2050, investments shift toward lithium-ion BESS, driven by the assumed future cost reductions (see \autoref{fig:CAPEX_storage}). Notably, the optimal storage capacity is 11.8\% higher in 2025 and 48.8\% higher in 2050 when reserve market participation is considered.

TES emerges as a consistently attractive option across all instances tested in this case study, supporting the findings of \cite{beck2021optimal_selection_of_storage}. TES investments show relatively limited sensitivity to reserve market participation, as anticipated, with an increase of 32.7\% with reserve market participation and 29.9\% without between 2025 and 2050. 

Among conversion technologies, electric and gas boilers dominate the 2025 investments. However, increasing CO$_2$ prices by 2050 drive a shift toward more sustainable alternatives. Electric boiler investments remain relatively stable, increasing by only 3.2\%, while gas boiler investments decreases by 10.2\%, reflecting reduced cost-effectiveness under the assumption of increasing emission costs. Heat pumps, on the other hand, become significantly more attractive by 2050, with investment increases of 64.4\% and 55.8\% with and without reserve market participation, respectively, across all temperature levels relevant for this technology. In parallel, PV enters the technology mix in 2050 - compared to no investments in 2025 - as a result of increasing emission costs, and reduced investment costs.

In summary, reserve market participation, increasing emission costs, and declining technology costs are key drivers of flexible asset investments for industrial prosumers. The optimal storage shifts from flywheel in 2025 to lithium-ion in 2050, reflecting the assumed future cost developments. TES remains attractive, with limited sensitivity to the reserve market. Increasing CO$_2$ prices along with decreasing investment costs reduce gas boiler investments and increase heat pumps and PV installation by 2050.

\section{Conclusion} \label{sec:conclusion}
A multi-stage stochastic programming model was utilized to analyze multi-market participation under uncertainty for a single industry setup. The results show that investment in flexible assets is beneficial both with and without reserve market participation. Therefore, industries already considering such investment should also optimize the storage size with reserve markets in mind. However, future work should expand to include a more detailed representation of the bidding sequences, as well as the inclusion of stand-alone activation markets, and multiple reserve markets to capture technology-specific qualification and bidding potential.

%% file: main.bbl
% Generated by IEEEtranN.bst, version: 1.14 (2015/08/26)